\documentclass{PoS}

\usepackage{amsfonts,amssymb,amsmath,bm}
\usepackage{graphicx}

\newcommand{\be}{\begin{equation}}
\newcommand{\bea}{\begin{eqnarray}}
\newcommand{\ee}{\end{equation}}
\newcommand{\eea}{\end{eqnarray}}
\newcommand{\bpi}{\begin{picture}}
\newcommand{\bce}{\begin{center}}
\newcommand{\epi}{\end{picture}}
\newcommand{\ece}{\end{center}}

\def\chic#1{{\scriptscriptstyle #1}}

\def\gb{\bm{\Gamma}}

\title{Non-perturbative Green's functions \\ and the QCD effective charge}

\ShortTitle{Non-perturbative Green's functions and the QCD effective charge}

\author{\speaker{Arlene~C.~Aguilar}\\
        Federal University of ABC, CCNH,\\ 
Rua Santa Ad\'elia 166,  CEP 09210-170, Santo Andr\'e, Brazil \\
        E-mail: \email{arlene.aguilar@ufabc.edu.br}}

\abstract{
Using as ingredients the non-perturbative  solutions of various  QCD 
Green's function obtained from Schwinger-Dyson  equations (SDEs), we study 
two  versions of  the QCD effective  charge. The first one  obtained 
from  the  pinch  technique  gluon self-energy,  and  the second from
the  ghost-gluon  vertex. Despite the distinct nature of their buildings blocks, the two effectives 
charges are almost identical in the entire range of momenta, due to a
fundamental  identity relating the 
ghost dressing function with the  two form factors of Green's function, which  is of central importance in the PT-BFM formalism. In this talk, we outline how to  
derive this crucial identity  from the SDEs 
of the aforementioned Green's functions. 
The renormalization 
procedure that preserves the validity of this identity is discussed in detail.
Most importantly, we show that  due to the infrared finiteness of  the gluon  propagator, 
the  QCD charge obtained  with either  definition  freezes in  the  deep infrared,  in
agreement  with theoretical  and  phenomenological expectations.}

\FullConference{International Workshop on QCD Greens Functions, Confinement, and Phenomenology - QCD-TNT09\\
		 September 07 - 11 2009\\
		 ECT Trento, Italy}

\begin{document}

\section{Introduction}

One of the most difficult problems in QCD is to understand the  interface  between the perturbative and non-perturbative regimes. Both sophisticated theoretical tools~\cite{Cornwall:1982zr,Binosi:2002vk}
 as well as more phenomenologically oriented approaches~\cite{Mattingly:1993ej} indicate that this connection is not abrupt, but rather smooth. 

Without any doubt, a continuous 
interpolation between the perturbative and  the non-perturbative region  is  
intimately  related to the  behavior of the  QCD fundamental coupling~\cite{Aguilar:2009nf}. However, we know by now,  
that the smoothness in this transition can not be achieved with perturbative assumption 
of a coupling with singular growth (Landau pole) in the infrared. Indeed,  all accumulated evidence points toward the need of a freezing of the QCD effective 
charge at small momenta \cite{Cornwall:1982zr,Binosi:2002vk,Mattingly:1993ej}, where $\alpha(q^2)$ develops an infrared fixed point and  QCD has a conformal window  at low energy~\cite{Brodsky:2003px}.

The infrared finiteness of the effective charge can be considered as one of   
the manifestation of the phenomenon of dynamical gluon mass 
generation~\cite{Cornwall:1982zr,Aguilar:2002tc} revealing
in this way, its  profound connection with the most fundamental Green's functions 
of QCD, such as the gluon and ghost propagators~\cite{Aguilar:2009nf}. Indeed, the basic ingredients that enter in its definition must contain the right information and be combined in a very precise way  in order to endow the effective charge with the required physical and 
field-theoretic properties~\cite{Binosi:2002vk}.

In this talk, we will show how different QCD Green's functions 
can be combined in order to form renormalization group (RG) invariant
quantities which eventually may be associated to  a definition of an  
effective charge~\cite{Aguilar:2009nf}. 
Specifically, we will 
consider firstly the effective charge obtained within the 
pinch technique (PT) framework~\cite{Cornwall:1982zr,Watson:1996fg}, and its correspondence~\cite{Binosi:2002ft} with the 
background-field method (BFM)~\cite{Abbott:1980hw}. The PT effective charge constitutes the 
most direct non-abelian generalization of the familiar concept of the 
QED effective charge. The  second definition involves the ghost and gluon self-energies~\cite{Alkofer:2004it}, in the Landau gauge, and in the kinematic configuration where the well-known Taylor non-renormalization theorem \cite{Taylor:1971ff} becomes applicable.  

\section{Definitions and ingredients}

Let us introduce some of the basic ingredients necessary for 
the definition of the two effective charges we want to study. In the Landau gauge, the full gluon propagator $\Delta_{\mu\nu}(q)$ is 
transverse, and omiting the color indices, 
its general form is given by
\be \Delta_{\mu\nu}(q)=-iP_{\mu\nu}(q)\Delta(q^2), \quad \mbox{with} \quad 
P_{\mu\nu}(q)= g_{\mu\nu} - \frac{q_\mu q_\nu}{ q^2}\,,
\label{prop_cov}
\ee
where the scalar function  $\Delta(q^2)$ is related 
to the all order self-energy  $\Pi_{\mu\nu}(q)=P_{\mu\nu}(q)\Pi(q^2)$ through
$\Delta^{-1}(q^2) = q^2 + i \Pi(q^2)$. The full ghost propagator $D(q^2)$ and its dressing function $F(q^2)$ are related by 
\be
D(q^2)= \frac{iF(q^2)}{q^2}.
\label{ghostdress}
\ee

\begin{figure}[ht]
\begin{center}
\includegraphics[scale=0.7]{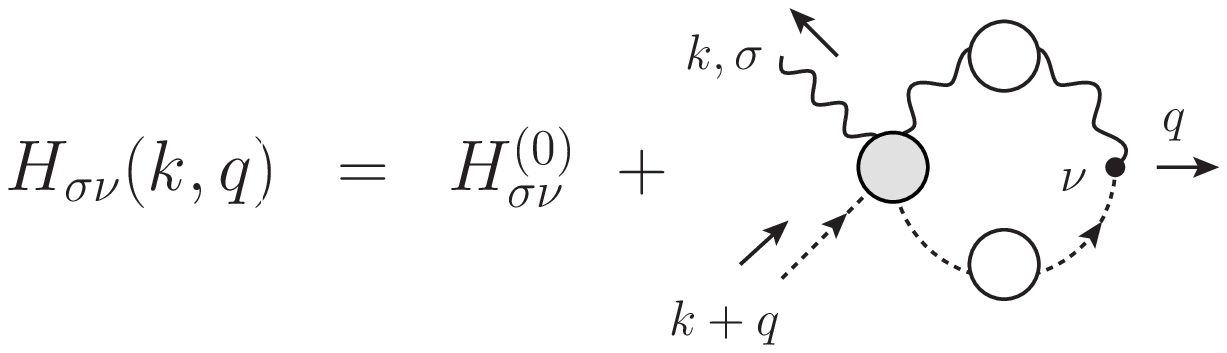}
\vspace{-0.3cm}
\caption{Diagrammatic representation of $H$.}
\label{fig:h}
\end{center}
\end{figure}

In our construction, a special role is played by the auxiliary two-point function $\Lambda_{\mu\nu}(q)$, represented in Fig.~\ref{fig:h} and  defined as~\cite{Aguilar:2008xm}  
\bea
 \Lambda_{\mu \nu}(q) = g_{\mu\nu} G(q^2) + \frac{q_{\mu}q_{\nu}}{q^2} L(q^2) &=& -i g^2C_A
\int_k H^{(0)}_{\mu\rho}
D(k+q)\Delta^{\rho\sigma}(k)\, H_{\sigma\nu}(k,q),
\label{LDec}
\eea
where $C_{\rm {A}}$ is the Casimir eigenvalue of the adjoint representation
[\,$\,C_{\rm {A}}\,=\,N$ for $SU(N)$\,], 
and $\int_{k}\equiv\mu^{2\varepsilon}(2\pi)^{-d}\int\!d^d k$, 
with $d=4-\epsilon$ the dimension of space-time. The vertex
$H_{\mu\nu}(k,q)$ is also represented in Fig.~\ref{fig:h}, and its 
tree-level counterpart is given \mbox{$H_{\mu\nu}^{(0)} = ig_{\mu\nu}$}.
An additional constraint on the behavior $H_{\mu\nu}(k,q)$  is imposed  by the
WI (Ward identity)
\be
q^\nu H_{\mu\nu}(k,q)=-i\gb_{\mu}(k,q)\,, 
\label{qH}
\ee 
where  $\gb_\mu(k,q)$  is the  all-order ghost vertex, with $k$ representing the 
momentum of the gluon and $q$ the one of the anti-ghost; at tree-level $\gb^{(0)}_\mu(k,q)=-q_\mu$.

\subsection{The pinch technique effective charge}

The heart of the PT effective charge definition 
lies on the construction of a new \emph{effective gluon propagator}, $\widehat{\Delta}(q^2)$,  
which captures the running of the QCD $\beta$ function, exactly as happens with the 
vacuum polarization in the case of QED 
\cite{Binosi:2002vk,Watson:1996fg,Binosi:2002ft}. Already at one-loop level, the PT gluon propagator displays the desired coefficient in 
front of the  perturbative logarithm, namely   
\be 
\widehat\Delta^{-1}(q^2)= q^2\left[1+ b g^2\ln\left(\frac{q^2}{\mu^2}\right)\right],
\label{rightRG}
\ee
where  $b = 11 C_A/48\pi^2$  is the first coefficient of the QCD $\beta$-function when the
number of fermions $n_f=0$ (quarkless QCD).  

\begin{figure}[ht]
\begin{center}
\includegraphics[scale=0.65]{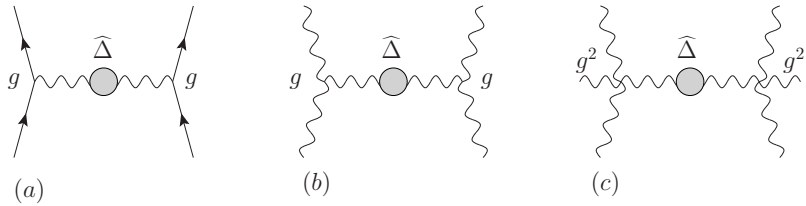}
\end{center}
\vspace{-0.7cm}
\caption{The universal PT coupling.}
\label{fig:pt_coup}
\end{figure}

In addition,  to all order $\widehat\Delta^{-1}(q^2)$ is universal (i.e. process-independent), and therefore it does not depend  on the details of the process where it is embedded as shown in Fig.~\ref{fig:pt_coup}. 

One important point, explained in detail in the literature, is the (all-order) correspondence 
between the PT and the Feynman gauge of the BFM~\cite{Binosi:2002ft,Abbott:1980hw}. 
In fact, one can generalize the PT construction~\cite{Binosi:2002ft}
in such a way as to reach diagrammatically 
any value of the gauge fixing parameter of the BFM, and in particular the Landau gauge.
In what follows we will implicitly assume the aforementioned generalization of the PT, 
given that the main identity we will use to 
relate the two effective charges is valid only in the Landau gauge.

Due to the Abelian WIs satisfied by the PT-
BFM Green's functions, the renormalization
constants of the gauge-coupling and of the PT
gluon self-energy, defined as
\bea
g(\mu^2) =Z_g^{-1}(\mu^2) g_0\,;\qquad
\widehat\Delta(q^2,\mu^2) =  \widehat{Z}^{-1}_A(\mu^2)\widehat{\Delta}_0(q^2), 
\label{conrendef}
\eea
where the ``0'' subscript indicates bare quantities, satisfy the 
QED-like relation 
\be
{Z}_{g} = {\widehat Z}^{-1/2}_{A}.
\label{ptwi}
\ee
Thus, it follows immediately that the product 
\be
{\widehat d}_0(q^2) = g^2_0 \widehat\Delta_0(q^2) = g^2(\mu^2) \widehat\Delta(q^2,\mu^2) = {\widehat d}(q^2), 
\label{ptrgi}
\ee
retains the same form before and after renormalization, i.e., it 
forms a RG-invariant ($\mu$-independent) quantity~\cite{Cornwall:1982zr,Binosi:2002vk,Binosi:2002ft}.
For asymptotically large momenta one may extract from ${\widehat d}(q^2)$
a dimensionless quantity by writing,
\be
{\widehat d}(q^2) = \frac{\overline{g}^2(q^2)}{q^2},
\label{ddef1}
\ee
where $\overline{g}^2(q^2)$ is the RG-invariant effective charge of QCD; at one-loop 
(use Eq.~(\ref{rightRG}) into (\ref{ptrgi}))
\be
\overline{g}^2(q^2) = \frac{g^2}{1+  b g^2\ln\left(q^2/\mu^2\right)}
= \frac{1}{b\ln\left(q^2/\Lambda^2_{\mathrm{QCD}}\right)} \,,
\label{effch}
\ee
where $\Lambda_{\mathrm{QCD}}$ denotes an RG-invariant mass scale of a few hundred ${\rm MeV}$.

Being a direct consequence of the WIs satisfied by the PT Green's function, Eq.~(\ref{ptrgi}) may be employed either perturbatively or non-perturbatively, provided that one has information on the IR behavior of the PT-BFM gluon propagator $\widehat\Delta(q^2)$.

However, thanks to a general relation connecting the PT-BFM $\widehat\Delta(q^2)$ and the {\it conventional} gluon propagator  $\Delta(q^2)$,  all the non-perturbative information we have gathered about $\Delta(q^2)$  may also be used.  
Specifically, the aforementioned  formal all-order relation 
states that~\cite{Grassi:1999tp}
\be
\Delta(q^2) = 
\left[1+G(q^2)\right]^2 \widehat{\Delta}(q^2), 
\label{bqi2}
\ee
where $G(q^2)$ is the form factor of  the $g_{\mu\nu}$ component appearing on the 
definition of  $\Lambda^{\mu\nu}$ given in Eq.(\ref{LDec}). Note that, due to its BRST origin, the above relation must be preserved after renormalization. 
Specifically, denoting by $Z_\Lambda$ the renormalization constant relating 
the bare and renormalized functions, $\Lambda_0^{\mu\nu}$ and $\Lambda^{\mu\nu}$, through
\be
g^{\mu\nu} + \Lambda^{\mu\nu}(q,\mu^2)=Z_\Lambda(\mu^2)[g^{\mu\nu}+ \Lambda_0^{\mu\nu}(q)],
\label{Lamrel}
\ee
then from (\ref{ptwi}) and (\ref{bqi2}) follows the additional relation 
\be
Z_g^{-1} = Z_A^{1/2} Z_\Lambda ,
\label{extrel}
\ee 
which it will be useful in the following subsection.

At lowest order, it is straightforward to  verify, that the role of the function 
the $1+G(q^2)$, obtained from Eq.~(\ref{LDec}), is to  restore the $\beta$ function coefficient  
in front of UV logarithm.  Explicitly, we have  at one-loop (in the Landau gauge)~\cite{Aguilar:2008xm}
\bea
1+G(q^2) = 1 +\frac{9}{4}
\frac{C_{\rm {A}}g^2}{48\pi^2}\ln\left(\frac{q^2}{\mu^2}\right)\; \qquad
\Delta^{-1}(q^2) &=& q^2 \left[1+\frac{13}{2}
\frac{C_{\rm {A}}g^2}{48\pi^2}\ln\left(\frac{q^2}{\mu^2}\right)\right].
\label{pert_gluon}
\eea
Using  Eq.~(\ref{bqi2}) we therefore recover the $\widehat{\Delta}^{-1}(q^2)$ 
of Eq.~(\ref{rightRG}), as we should. 

Then, non-perturbatively, one substitutes into  Eq.~(\ref{bqi2}) the $ 1+G(q^2)$ and $\Delta(q^2)$ 
obtained from either the lattice or SD analysis, to obtain $\widehat{\Delta}(q^2)$.  
As explained above, the combination formed by  
\be
\widehat{d}(q^2)= \frac{g^2 \Delta(q^2)}{\left[1+G(q^2)\right]^2} \,,
\label{rgi}
\ee
is independent of the renormalization point $\mu$  {\it i.e.} a RG-invariant quantity.

\subsection{Gluon-ghost vertex}

Another possibility for defining the QCD effective charge can be
obtained starting  from the various QCD vertices. The basic idea behind
is to recognize the RG-invariant quantities we may form out of these
vertices. The downside of this construction lies in the fact that it   
involves all the momentum scales present in the vertex in question, 
and further assumptions about their kinematic configuration need  to be introduced, in order to express the charge as a function of a single variable. 
The ghost-gluon vertex has been particularly popular in this context, 
especially in conjunction with Taylor's non-renormalization theorem and 
the corresponding kinematics~\cite{Alkofer:2004it}.
\begin{figure}[ht]
\begin{center}
\includegraphics[scale=0.65]{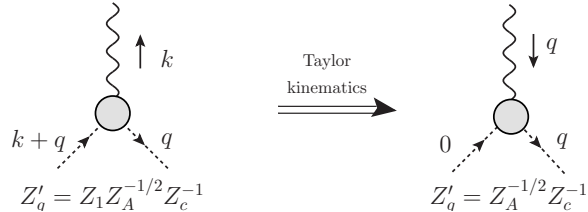}
\end{center}
\vspace{-0.65cm}
\caption{The ghost-gluon vertex and the Taylor kinematics.}
\label{fig:gvr}
\end{figure}

For the case of the ghost-gluon vertex (see Fig.~\ref{fig:gvr}), the renormalization constants involved are
\bea
\Delta(q^2,\mu^2)= Z^{-1}_{A}(\mu^2)\Delta_0(q^2), &\qquad&
F(q^2,\mu^2)= Z^{-1}_{c}(\mu^2)F_0(q^2),\nonumber\\ 
\gb^\nu(k,q,\mu^2)= Z_1(\mu^2)\gb^\nu_0(k,q),&\qquad& 
g_0= Z_{g^{\prime}}(\mu^2)g^{\prime}.
\label{renconst}
\eea

Notice that a priori $Z_{g^{\prime}}$ defined as
$Z_{g'}= Z_1 Z_A^{-1/2} Z_c^{-1}$,
does not have to coincide with the 
$Z_{g}$ appearing in (\ref{conrendef}); however, as we will see 
in the next section, they do coincide by virtue of the basic identity 
we will derive there.

Choosing the special Taylor's kinematic configuration, where the incoming ghost
has a vanishing momentum (i.e. $k_\mu \to -q_\mu$), one may impose the following additional condition one the renormalization constant  $Z_1$ (valid only in the Landau gauge), namely  
$Z_1= Z_{g^{\prime}} Z_A^{1/2}Z_c=1$, from which follows immediately that 
\be
Z_{g^{\prime}}^{-1} = Z_A^{1/2}Z_c. 
\label{mores}
\ee
Thus, the product
\be
\widehat{r}(q^2)  =  {g'}^2 \Delta(q^2;\mu^2) F^2(q^2;\mu^2) 
 = {g'}_0^2 \Delta_0(q^2) F^2_0(q^2),
\label{rg2}
\ee
forms a dimensionful $\mu$-independent combination. Therefore, for asymptotically large $q^2$, in analogy to Eq.~(\ref{ddef1}), one can define  
an alternative QCD running coupling as 
\be
\widehat{r}(q^2)=\frac{\overline{g}_{\mathrm{gh}}^2(q^2)}{q^2}.
\ee 
Using then Eq.~(\ref{pert_gluon}), and the fact that
\be
F^{-1}(q^2) = 1+\frac94\frac{C_{\rm {A}}g^2}{48\pi}\ln\left(\frac{q^2}{\mu^2}\right),
\ee
it is straightforward to verify that  $\overline{g}_\mathrm{gh}^2(q^2)$ and  $\overline{g}^2(q^2)$ display the same one-loop 
behavior, since, perturbatively the function $1+G(q^2)$ is the inverse of the ghost dressing function $F(q^2)$.
As we will see in the next section, this is nothing more than the one-loop manifestation 
of the more general identity relating $G(q^2)$ and $F(q^2)$.

\section{\label{giSDE} A special relation between Green's functions}

In this section, we sketch the  main steps needed to derive the central identity, valid  {\it only} in the Landau gauge, relating the ghost dressing function with 
a particular combination of the form-factors $G(q^2)$ and $L(q^2)$ 
appearing in the tensorial decomposition of $\Lambda_{\mu\nu}$  given in Eq.~(\ref{LDec}).  
\begin{figure}[ht]
\begin{center}
\includegraphics[width=11cm]{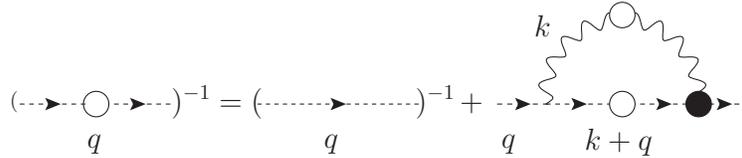}
\end{center}
\vspace{-0.6cm}
\caption{The SDE for the ghost.}
\label{fig:ghostSDE}
\end{figure}

First, consider the standard SD equation for the ghost propagator (Fig~\ref{fig:ghostSDE}), 
\be
iD^{-1}(q^2) = q^2 +i g^2 C_{\rm {A}}  \int_k
\Gamma^{\mu}\Delta_{\mu\nu}(k)\gb^{\nu}(k,q) D(k+q).
\label{SDgh}
\ee
Then, contract both sides of
the defining equation~(\ref{LDec}) by the combination $q^{\mu}q^{\nu}$ to get
\be
[G(q^2) + L(q^2)]q^2 = g^2 C_{\rm {A}}
\int_k q_{\rho} \Delta^{\rho\sigma}(k)\, q^{\nu} H_{\sigma\nu}(k,q) D(k+q).
\label{s1}
\ee
Using Eq.~(\ref{qH}) and the transversality of the full gluon propagator, we can see that the rhs of Eq.~(\ref{s1}) 
is precisely the integral appearing in the ghost SDE~(\ref{SDgh}). Therefore 
\be
[G(q^2) + L(q^2)]q^2 =  iD^{-1}(q^2) - q^2,
\label{s2}
\ee
or, in terms of the ghost dressing function $F(q^2)$ [{\it viz}. Eq.~(\ref{ghostdress})]  
\be
\boxed{1+ G(q^2) + L(q^2) = F^{-1}(q^2)}
\label{funrel}
\ee

The above relation, derived here from the SD equations of the theory,  
has been first obtained in~\cite{Kugo:1995km}, in the framework of the Batalin-Vilkovisky quantization formalism. As was shown there, the relation is a direct consequence of the fundamental BRST symmetry.

Now, let us construct the dynamical equations governing the behavior of the
functions  $G(q^2)$ and  $L(q^2)$. The tensorial projection of 
both functions in terms of $\Lambda_{\mu\nu}$ may be obtained from Eq.~(\ref{LDec}), where 
in $d$ dimensions, we have 
\be
G(q^2) = \frac{1}{(d-1)q^2} \left(q^2 \Lambda_{\mu}^{\mu} - q^{\mu}q^{\nu}\Lambda_{\mu\nu} 
\right),\qquad L(q^2) = \frac{1}{(d-1)q^2} \left(d q^{\mu}q^{\nu}\Lambda_{\mu\nu} - q^2\Lambda_{\mu}^{\mu} \right),
\label{s3}
\ee 
which then gives, in terms of the SDE integrals
\bea
G(q^2) &=& \frac{g^2 C_{\rm {A}}}{d-1}
\left[ 
\int_k \Delta^{\rho\sigma}(k)\, H_{\sigma\rho}(k,q) D(k+q)
+i
\frac{1}{q^2} \int_k q^{\rho} \Delta_{\rho\sigma}(k)\, {\gb}^{\sigma}(k,q) D(k+q)
\right],
\nonumber\\
L(q^2) &=& -\frac{g^2 C_{\rm {A}}}{d-1}
\left[i
\frac{d}{q^2} \int_k q^{\rho} \Delta_{\rho\sigma}(k)\, {\gb}^{\sigma}(k,q) D(k+q)
\!+\!\! \int_k \Delta^{\rho\sigma}(k)\, H_{\sigma\rho}(k,q) D(k+q)
\right]\!.
\label{s4}
\eea 

In this point some words about the renormalization and approximations we will employ in  Eq.~(\ref{s4}) are in order. Let us start with the renormalization procedure.
As mentioned before,  since the origin of (\ref{funrel}) 
is the BRST symmetry,  it should not be deformed after renormalization.
Combining the definitions of (\ref{Lamrel}) and  (\ref{renconst}), we see that
in order to preserve the relation~(\ref{funrel}) we must impose that 
\mbox{$Z_\Lambda = Z_{c}$}.
In addition, by virtue of (\ref{qH}), and for the same reason,  
we have that, in the Landau gauge, ${\gb}_{\nu}(k,q)$ and $H_{\sigma\nu}(k,q)$ 
must be renormalized by the same renormalization constant, namely $Z_1$ 
[{\it viz.} Eq.~(\ref{renconst})]; for the Taylor kinematics, we have that $Z_1=1$.  
  
Then, approximating the two vertices, \mbox{$H_{\mu\nu}(k,q)=ig_{\mu\nu}$}, and \mbox{$\gb_{\mu}(k,q) = -q_{\mu}$}, by their tree-level values, then, setting \mbox{$f(k,q) \equiv  (k \cdot q)^2/{k^2 q^2}$}, one may show that \cite{Aguilar:2009nf} 
\bea
F^{-1}(q^2) &=& Z_c +g^2 C_{\rm {A}} \int_k [1-f(k,q)] \Delta (k)  D(k+q),
\nonumber\\
1+G(q^2) &=& Z_c + \frac{g^2 C_{\rm {A}}}{d-1}\int_k [
(d-2)+ f(k,q)]\Delta (k)  D(k+q),
\nonumber\\
L(q^2) &=& \frac{g^2 C_{\rm {A}}}{d-1}\int_k 
[1 - d \,f(k,q)]\Delta (k)  D(k+q)\,,
\label{simple}
\eea 
which clearly satisfies Eq.~(\ref{funrel}).

We next go to the Euclidean space, by 
setting $-q^2=q^2_\mathrm{E}$, and defining $\Delta_\mathrm{E}(q^2_\mathrm{E})=-\Delta(-q^2_\mathrm{E})$, $D_\mathrm{E}(q^2_\mathrm{E})=-D(-q^2_\mathrm{E})$, and for the integration measure $\int_k=i\int_{k_\mathrm{E}}$. Then, suppressing the subscript ``E'' and  setting $q^2=x$, $k^2=y$,  we have that \mbox{$k \cdot q =\sqrt{xy}\cos\theta$}, and so $(k \cdot q)^2/q^2 =y\cos^2\theta $, and 
$(k+q)^2 = x + y +2 \sqrt{xy}\cos\theta$, we arrive at (see details in \cite{Aguilar:2009nf})
\bea
1+G(x) &=&  Z_c - \frac{\alpha_s C_{\rm {A}}}{16\pi}\left[
\frac{F(x)}{x}\int_{0}^{x}\!\!\! dy\  y \left(3 + \frac{y}{3x}\right) \Delta(y) 
+ \int_{x}^{\infty}\!\!\! dy \left(3 + \frac{x}{3y}\right)\Delta(y)F(y) 
\right],
\nonumber\\
L(x) &=&  \frac{\alpha_s C_{\rm {A}}}{12\pi} \left[
\frac{F(x)}{x^2}\int_{0}^{x}\!\!\! dy\ y^2 \Delta(y) 
+ x \int_{x}^{\infty}\!\!\! dy \frac{\Delta(y) F(y)}{y}
\right],
\nonumber\\
F^{-1}(x) &=& Z_c - \frac{\alpha_s C_{\rm {A}}}{16\pi} \left[
\frac{F(x)}{x}\int_{0}^{x}\!\!\! dy\  y \left(3 - \frac{y}{x}\right) \Delta(y) 
+ \int_{x}^{\infty}\!\!\! dy \left(3 - \frac{x}{y}\right)\Delta(y) F(y) 
\right].
\label{LFGr}
\eea

Then, it is  easy to see ({\it e.g.}, by means of the change of variables $y=zx$) that if $\Delta$ and $F$ are IR finite, Eq.~({\ref{LFGr}}) yields the important result $L(0)=0$ \cite{Aguilar:2009nf}.
Let us now assume that the renormalization condition for $F(x)$ was chosen 
to be \mbox{$F(\mu^2) =1$}. This condition, when inserted into the third equation of (\ref{LFGr}), 
allows one to express $Z_c$ as  
\be
Z_c = 1+ \frac{\alpha_s C_{\rm {A}}}{16\pi}  \left[
\frac{1}{\mu^2}\int_{0}^{\mu^2}\!\!\! dy  y \left(3 - \frac{y}{\mu^2}\right) \Delta(y) 
+ \int_{\mu^2}^{\infty}\!\!\! dy \left(3 - \frac{\mu^2}{y}\right)\Delta(y) F(y) 
\right],
\label{Zcexp}
\ee
and may be used to cast (\ref{LFGr}) into a manifestly renormalized form.
Note that if one choses \mbox{$F(\mu^2) =1$}, then one cannot choose simultaneously  
\mbox{$G(\mu^2)=0$}, because that would violate the identity of Eq.~(\ref{funrel}), 
given that \mbox{$L(\mu^2)\neq 0$}. In fact, once $F(\mu^2) =1$ has been imposed, 
the value of $G(\mu^2)$ is completely determined from its own equation, i.e. 
the first equation in~(\ref{LFGr}). 

In addition, in the MOM scheme  $\Delta(q^2)$ and $\widehat{\Delta}(q^2)$ cannot be made equal at the renormalization point, since  Eq.~(\ref{bqi2})  implies
$\widehat{\Delta}^{(-1)}(\mu^2) = \mu^{2}\left[ 1+G^2(\mu^2) \right]^{2}$.

Now, let us to  return to the couplings, and discuss the implications
of the identity given by Eq.~(\ref{funrel}). First of all, comparing Eq.~(\ref{extrel}) and Eq.~(\ref{mores}), it is clear that $g(\mu)=g^{\prime}(\mu)$, by virtue of \mbox{$Z_\Lambda = Z_{c}$}. Therefore, using Eq.~(\ref{bqi2}), and the definitions given in Eqs.~(\ref{ptrgi}) and (\ref{rg2}), one can obtain a relation between the two RG-invariant quantities, 
$\widehat{r}(q^2)$ and $\widehat d(q^2)$, namely
\be
\widehat{r}(q^2)=[1+G(q^2)]^2 F^2(q^2)\widehat d(q^2).
\label{rel_rgi}
\ee
From this last equality, follows that $\alpha_{\chic{\mathrm{PT}}}(q^2)$ and $\alpha_{\mathrm{gh}}(q^2)$ 
are related by 
\be
\alpha_{\mathrm{gh}}(q^2) = [1+G(q^2)]^2 F^2(q^2)\alpha_{\chic{\mathrm{PT}}}(q^2),
\label{coup_charge}
\ee

After using Eq.~(\ref{funrel}), we have that  

\be
\alpha_{\chic{\mathrm{PT}}}(q^2) = \alpha_{\mathrm{gh}}(q^2)\left[1+ \frac{L(q^2)}{1+G(q^2)}\right]^2 \,.
\label{relcoup2}
\ee
Evidently, the two couplings can only coincide at two points:
(i) at $q^2=0$, where, 
due to the fact that $L(0)=0$ , 
we have that 
\mbox{$\alpha_{\mathrm{gh}}(0) = \alpha_{\chic{\mathrm{PT}}}(0)$}, 
and (ii) at \mbox{$q^2= \infty$},
given that in the deep UV $L(q^2)$ approaches a constant.
Note in fact that  
the two effective charges {\it cannot} coincide 
at the renormalization point $\mu$, where  
\mbox{$\alpha_{\mathrm{gh}}(\mu^2) = [1-L(\mu^2)]^2 \alpha_{\chic{\mathrm{PT}}}(\mu^2)$};
this can be understood also in terms of the discussion following Eq.~(\ref{Zcexp}). 

\section{Schwinger-Dyson input and numerical analysis}

Now we are in the position to compute the QCD effective charges defined above, 
using as input the non-perturbative solutions of SDE for the various Green's functions appearing in their definitions. More specifically, we will solve numerically a system of three coupled non-linear integral equations in the Landau gauge, 
containing $\Delta(q^2)$, $F(q^2)$, and $G(q^2)$ as unknown quantities. 

Once solutions for these three functions have been obtained, then $L(q^2)$
is fully determined by its corresponding equation, namely the second one in Eq.~(\ref{LFGr}). 
\begin{figure}[ht]
\includegraphics[width=15cm]{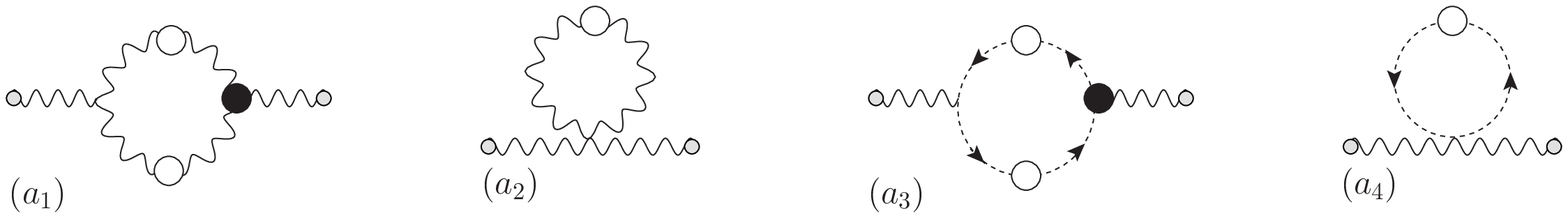}
\caption{The new SDE for the gluon one-loop dressed diagrams.}
\label{fig:SDeqs}
\end{figure}

The two SD equations determining  $F(q^2)$, $G(q^2)$ are given in Eq.~(\ref{LFGr}).
The SD equation governing $\Delta(q^2)$, is given by~\cite{Aguilar:2008xm} 
\be
[1+G(q^2)]^2\Delta^{-1}(q^2)P_{\mu\nu}(q) = 
q^2 P_{\mu\nu}(q) + i\sum_{i=1}^{4}(a_i)_{\mu\nu},
\label{SDgl}
\ee
where the diagrams $(a_i)_{\mu\nu}$ are shown in Fig.~\ref{fig:SDeqs}. As explained in~\cite{Aguilar:2008xm}, 
due to the abelian WI satisfied by the fully-dressed vertices in the PT-BFM scheme, 
we have that  $q^{\mu}[(a_1)_{\mu\nu}+ (a_2)_{\mu\nu}] = q^{\mu}[(a_3)_{\mu\nu}+ (a_4)_{\mu\nu}] =0$.
This last property enforces the 
the transversality of the gluon self-energy ``order-by-order'' in the dressed-loop expansion, 
which is one of the central features of the gauge-invariant 
SD truncation scheme defined within the PT-BFM framework. 

After introducing appropriate Ans\"atze for the aforementioned  fully-dressed vertices, 
we finally arrive at the integral equation
\bea
[1+G(q^2)]^2 \Delta^{-1}(q^2)\, &=&\, 
q^2 - \frac{g^2C_A}{6} \left[ 
\int_k \!\!\Delta(k)\Delta(k+q)f_1 
+ \int_k  \!\!\Delta(k) f_2 
- \frac{1}{2} \int_k \frac{q^2}{k^2 (k+q)^2}\right] 
\nonumber\\
&+&  g^2 C_A \bigg[\frac{4}{3}
\int_k \left[ k^2 - \frac{(k\cdot q)^2}{q^2}\right] D(k) D(k+q)
- 2 \int_k  D(k)\bigg]\,,
\label{sdef}
\eea
with
\bea
f_1 &=& 20q^2 + 18k^2 -6(k+q)^2 + \frac{(q^2)^2}{(k+q)^2}-  (k\cdot q)^2\bigg[ \frac{20}{k^2}
+ \frac{10}{q^2} + \frac{q^2}{k^2 (k+q)^2}
+\frac{2 (k+q)^2}{q^2 k^2}\bigg] \,,\nonumber\\
f_2 &=& -\frac{27}{2} -8 \frac{ k^2}{(k+q)^2}
+8 \frac{q^2}{(k+q)^2} 
+ 4 \frac{(k\cdot q)^2}{k^2(k+q)^2}
- 4 \frac{(k\cdot q)^2}{q^2(k+q)^2} \,,\
\label{f1f2}
\eea
The important point is that, by virtue of the poles introduced into the equation through the 
particular Ans\"atze employed \cite{Cornwall:1982zr,Aguilar:2008xm,Jackiw:1973tr}, 
one obtains an IR finite solution for the gluon propagator, i.e.a solution with  
\mbox{$\Delta^{-1}(0) > 0$}, 
in complete agreement with a large body of lattice data~\cite{Cucchieri:2007md}. 

\begin{figure}[ht]
\begin{minipage}[b]{0.5\linewidth}
\centering
\hspace{-0.75cm}
\includegraphics[scale=0.75]{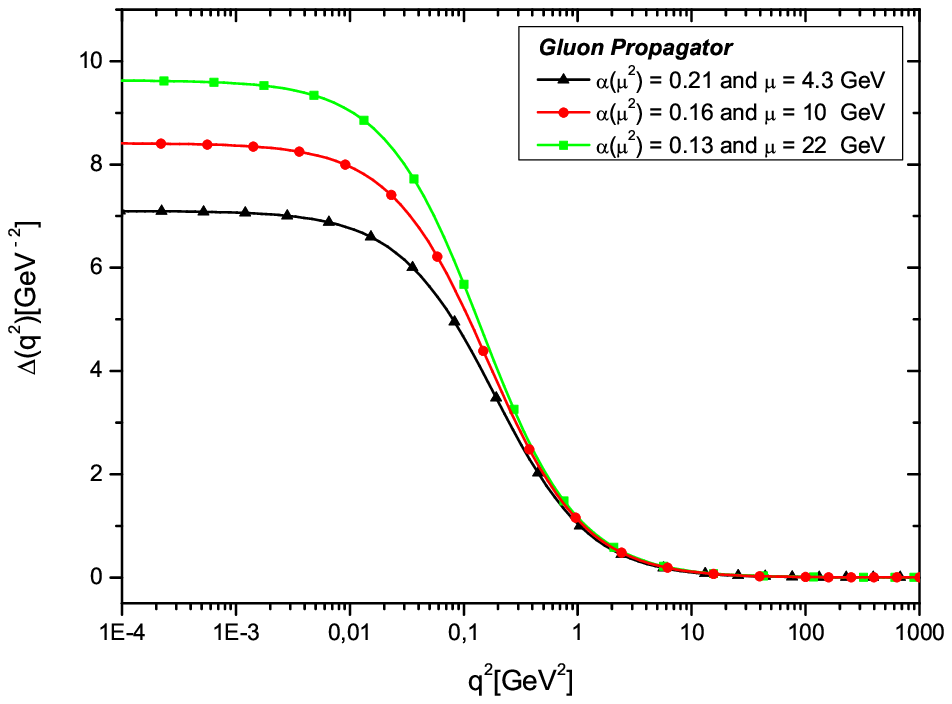}
\end{minipage}
\hspace{-0.5cm}
\begin{minipage}[b]{0.4\linewidth}
\centering
\hspace{-1.55cm}
\includegraphics[scale=0.75]{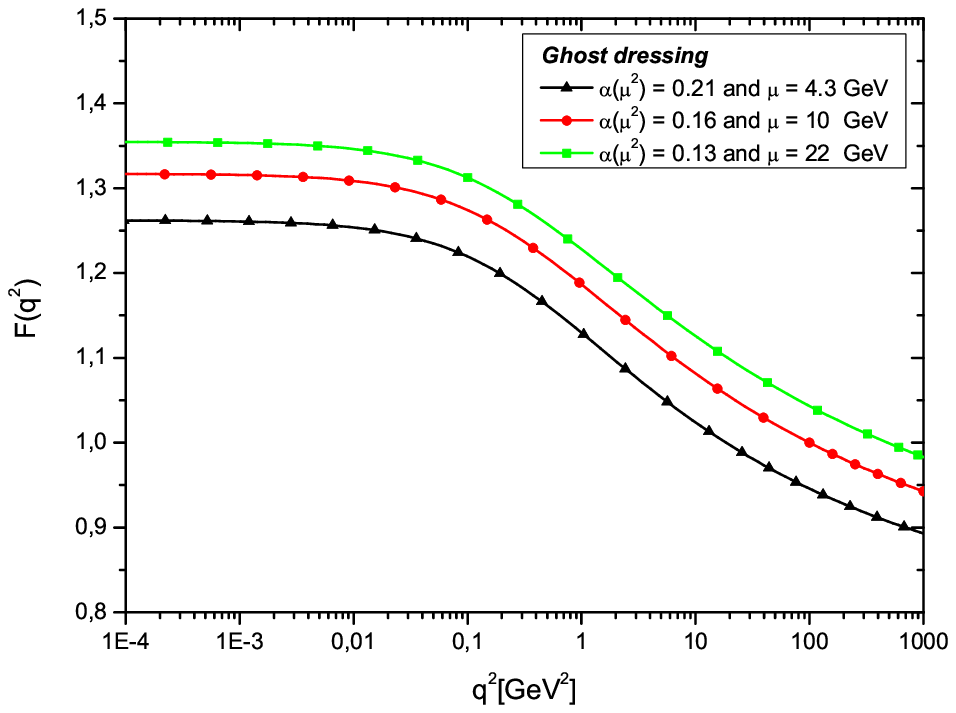}
\end{minipage}
\vspace{-0.5cm}
\caption{{\it Left panel}: Numerical solutions for the gluon propagator obtained from the SDE
renormalized at three different points, \mbox{$\mu = \{4.3, 10, 22\}$}\,\mbox{GeV} with
\mbox{$\alpha(\mu^2)=\{0.21, 0.16,0.13\}$}.
{\it Right panel}: The ghost dressing function $F(q^2)$ obtained from its corresponding SDE and renormalized at the same points.}
\label{fig:plot1}
\end{figure}

In Figs.~\ref{fig:plot1} we show the results for $\Delta(q^2)$ and $F(q^2)$ renormalized at three different points, \mbox{$\mu = \{4.3, 10, 22\}$} \mbox{GeV} with
  \mbox{$\alpha(\mu^2)=\{0.21, 0.16,0.13\}$} respectively. On the right panel we plot the corresponding $F(q^2)$ renormalized at the same points.  Notice that the solutions obtained are in qualitative agreement with recent results from large-volume lattices \cite{Cucchieri:2007md} where the both quantities, $\Delta(q^2)$ and $F(q^2)$, reach finite (non-vanishing) values in the deep IR.  
  
\begin{figure}[ht]
\begin{minipage}[b]{0.5\linewidth}
\centering
\hspace{-0.75cm}
\includegraphics[scale=0.75]{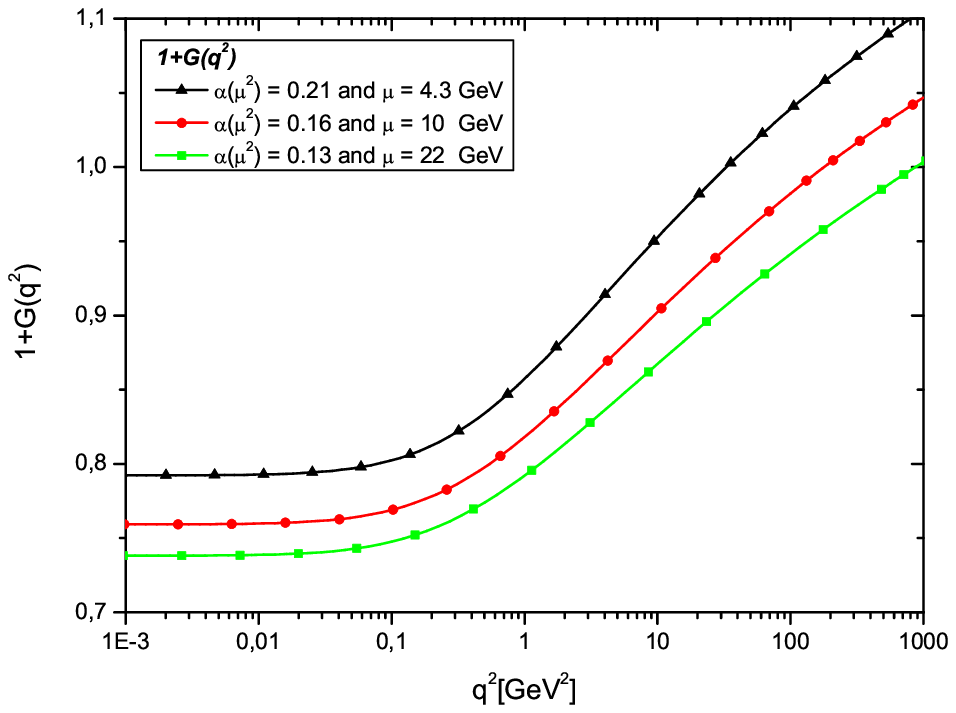}
\end{minipage}
\hspace{-0.5cm}
\begin{minipage}[b]{0.4\linewidth}
\centering
\hspace{-1.55cm}
\includegraphics[scale=0.75]{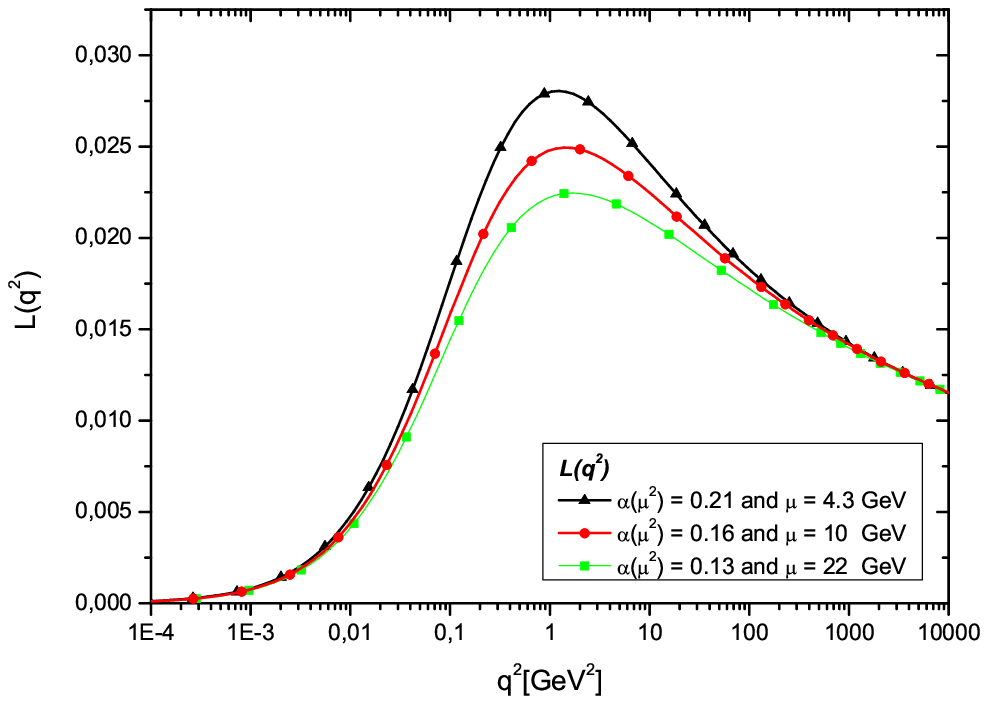}
\end{minipage}
\vspace{-0.5cm}
\caption{{\it Left panel}: \mbox{$1+G(q^2)$} determined from Eq.~(3.8), using the %
solutions for $\Delta(q^2)$ and $D(q^2)$ presented in the Fig.~6 at the same renormalization points. {\it Right panel}: The function $L(q^2)$ obtained from Eq.~(3.8).}
\label{fig:plot2}
\end{figure}

The results  for $1+G(q^2)$ and $L(q^2)$, renormalized at the same points, are presented in Fig.~\ref{fig:plot2}. As we can see, the function $1+G(q^2)$ is also IR finite exhibiting a plateau for  values of \mbox{$q^2<0.1 \mbox{GeV}^2$}. In the UV region, we instead recover the perturbative behavior~(\ref{pert_gluon}). On the other hand, $L(q^2)$ shows a maximum in the intermediate momentum region, while, as expected, $L(0)=0$.

\begin{figure}[ht]
\begin{minipage}[b]{0.5\linewidth}
\centering
\hspace{-0.75cm}
\includegraphics[scale=0.75]{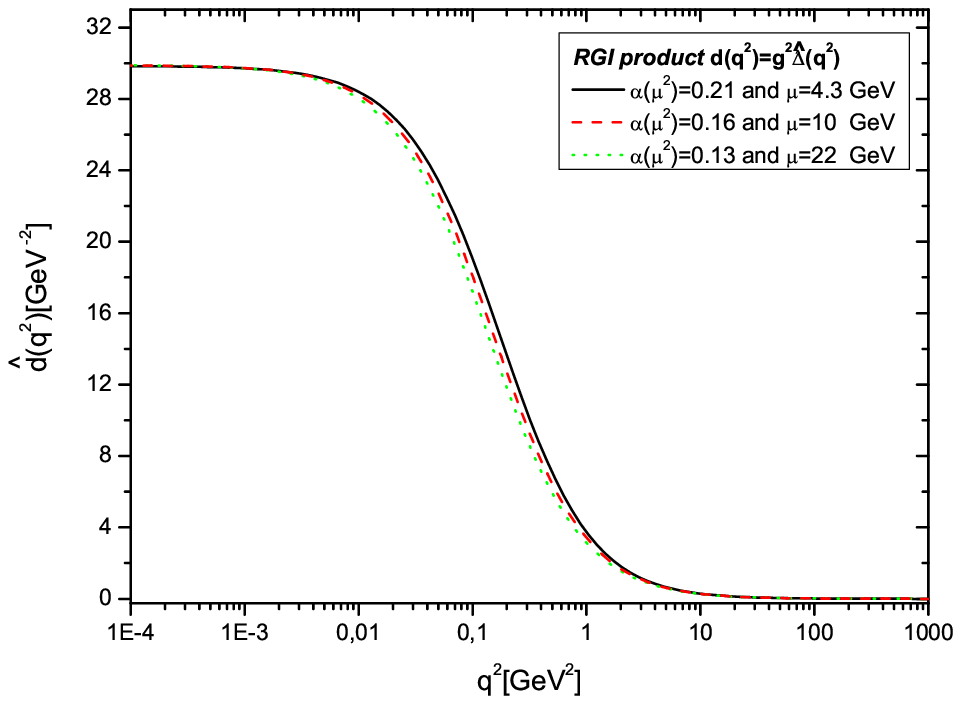}
\end{minipage}
\hspace{0.5cm}
\begin{minipage}[b]{0.45\linewidth}
\centering
\hspace{-0.75cm}
\includegraphics[scale=0.75]{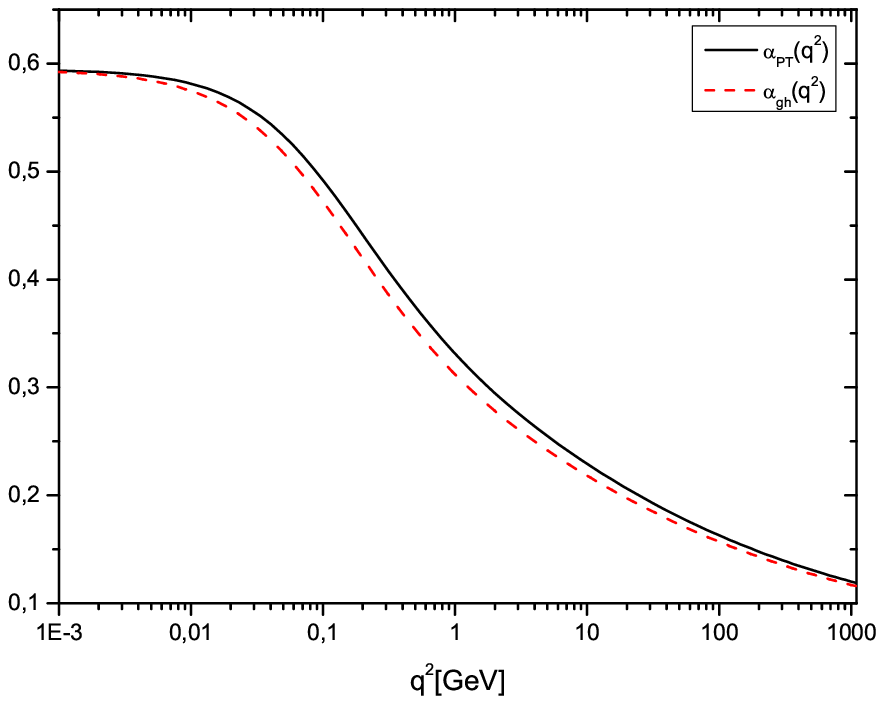}
\end{minipage}
\vspace{-0.65cm}
\caption{{\it Left panel}: The $\widehat{d}(q^2)$ obtained by combining $\Delta(q^2)$ and $[1+G(q^2)]^2$ according to Eq.~(2.15).
{\it Right panel}: $\alpha_{\mathrm{gh}}(q^2)$ vs 
$\alpha_{\mathrm{PT}}(q^2)$, for \mbox{$m_0=500 \,\mbox{MeV}$}.}
\label{fig:plot3}
\end{figure}

With all ingredients defined, the first thing one can check is that indeed Eq.~(\ref{rgi}) 
gives rise to a RG-invariant combination. Using the latter definition,  we can combine  
the different data sets  for $\Delta(q^2)$ and $[1+G(q^2)]^2$ at different renormalization points, to arrive at  the curves shown on  the left panel of Fig.~\ref{fig:plot3}.  Indeed, we see that all curves, for different values of $\mu$, merge one into the other proving 
that the combination $\widehat{d}(q^2)$ is independent of the renormalization point chosen.

From the dimensionful ${\widehat d}(q^2)$ we must now extract 
a dimensionless factor, ${\overline g}^2(q^2)$, corresponding to the running coupling (effective charge).  
Given that $\Delta(q^2)$ is IR finite, 
the physically meaningful procedure is to factor out from 
${\widehat d}(q^2)$ a massive propagator $[q^2+m^2(q^2)]^{-1}$, 
\be
\widehat{d}(q^2) = \frac{\overline{g}^2(q^2)}{q^2 + m^2(q^2)}\,,
\label{ddef}
\ee
where for the mass we will assume  ``power-law'' running~\cite{Lavelle:1991ve}, 
$m^2(q^2)= m^4_0/(q^2+m^2_0)$.

Thus, it follows from Eq.~(\ref{ddef}), that the effective charge $\alpha_{\chic{\mathrm{PT}}}(q^2) = \overline{g}^2(q^2)/4\pi$ is identified as being
\be
4\pi\alpha_{\chic{\mathrm{PT}}}(q^2)= [q^2+m^2(q^2)]\widehat{d}(q^2)\,.
\label{charge}
\ee 

Finally we compare numerically the two effective charges, $\alpha_{\chic{\mathrm{PT}}}(q^2)$ and $\alpha_{\mathrm{gh}}(q^2)$ on the right panel of Fig.~\ref{fig:plot3}. First, we determine $\alpha_{\chic{\mathrm{PT}}}(q^2)$ obtained using (\ref{charge}), then we obtain $\alpha_{\mathrm{gh}}(q^2)$ with help of (\ref{relcoup2}) and the results for $1+G(q^2)$ and $L(q^2)$ shown in Fig.~\ref{fig:plot2}.
As we can clearly see, both couplings freeze  at the same  finite value,
exhibiting a plateau for values of $q^2<0.02\, \mbox{GeV}^2$,  
while in the UV both show the expected  perturbative behavior. They differ only slightly in the intermediate region where
the values of $L(q^2)$ are appreciable.

\section{Conclusions}

In  this talk we have compared the definition of two QCD 
effective charges, $\alpha_\chic{{\mathrm{PT}}}(q^2)$ and $\alpha_{\mathrm{gh}}(q^2)$, 
obtained within two vastly different 
frameworks: the PT-BFM on the one hand, 
and the ghost-gluon vertex (with the Taylor-kinematics) on the other.

Despite their  distinct field-theoretic
origin, their dynamics involves the gluon propagator $\Delta(q^2)$ 
as a common ingredient and two different ingredients, 
which participate in a non-trivial identity. 
This identity, which is valid only in the Landau gauge, 
relates the ghost dressing function, $F(q^2)$, with the 
two form-factors,  $G(q^2)$ and $L(q^2)$. 

As consequence of the aforementioned identity, we have shown that the two effective charges are  almost identical in the entire range of physical  momenta.  More specifically, they coincide  exactly in  the deep  infrared, where
they  freeze at  a common  finite value, signaling the appearance of IR 
fixed point and a conformal window in QCD~\cite{Brodsky:2003px}, in agreement with
a  variety of phenomenological studies~\cite{Halzen:1992vd}.

\acknowledgments
I would like to thank the ECT* for the hospitality and for supporting
the QCD-TNT organization. This work was supported by Funda\c{c}\~ao de Amparo \`a Pesquisa do Estado de S\~ao Paulo (Fapesp)
under grant 2009/08721-3.

\end{document}